\documentclass[journal=jacsat,manuscript=article]{achemso}
\usepackage[version=3]{mhchem} 
\usepackage{hyperref}
\usepackage{xcolor}

\author{Addhyaya Sharma}
\affiliation{Department of Physics, Center for Discovery and Innovation, The City College of New York, 85 St. Nicholas Terrace, New York, NY 10031, USA.}
\author{Ezra Bader}
\affiliation{Department of Physics, Duke University, 1328 Campus Drive, Durham, NC 27708, USA.}
\author{Ravindra K. Yadav}
\affiliation{Department of Physics, Center for Discovery and Innovation, The City College of New York, 85 St. Nicholas Terrace, New York, NY 10031, USA.}
\altaffiliation{School of Physical Sciences, Indian Institute of Technology Mandi, Mandi, Himachal Pradesh 175005, India.}
\author{Juan Carlos Obeso Jureidini}
\affiliation{Department of Chemistry and Biochemistry, University of California San Diego, 9500 Gilman Drive, La Jolla, CA 92093, USA.}
\author{Michael Reitz}
\affiliation{Department of Chemistry and Biochemistry, University of California San Diego, 9500 Gilman Drive, La Jolla, CA 92093, USA.}
\author{Daegwang Choi}
\affiliation{Department of Physics, Center for Discovery and Innovation, The City College of New York, 85 St. Nicholas Terrace, New York, NY 10031, USA.}
\author{Rishabh Kaurav}
\affiliation{Department of Physics, Center for Discovery and Innovation, The City College of New York, 85 St. Nicholas Terrace, New York,,  NY 10031, USA.}
\altaffiliation{Ph.D. Program in Chemistry, The Graduate Center of CUNY, New York, NY 10016, USA.}
\author{Joel Yuen-Zhou}
\affiliation{Department of Chemistry and Biochemistry, University of California San Diego, 9500 Gilman Drive, La Jolla, CA 92093, USA.}
\author{Vinod M. Menon}
\affiliation{Department of Physics, Center for Discovery and Innovation, The City College of New York, 85 St. Nicholas Terrace, New York, NY 10031, USA.}
\altaffiliation{Ph.D. Program in Physics, The Graduate Center of CUNY, New York, NY 10016, USA.}
\email{vmenon@ccny.cuny.edu}

\title{Towards coherent polaritonic circuits operating at room temperature}

\begin{document}

\begin{abstract}

Polariton condensation is a potential system state for performing analog computations, given that it exhibits quantum behavior at macroscopic scales readily probed with low-cost optical methods. Current methods of fabricating devices in polariton microcavities largely involve patterning the devices via e-beam lithography before the cavity is completed, which offers less flexibility in device creation and reduces the maximum possible refractive index contrast. Moreover, the momentum and spatial distributions of the condensate are highly dependent on the host platform, and it has been difficult to preserve the desired behavior when modifying a given cavity. Here we introduce a method that addresses both of these challenges with the creation of polaritonic circuits of arbitrary forms etched via Focused Ion Beam into an organic microcavity based on Rhodamine 3B Perchlorate within a Small Molecule Ionic Isolation Lattices complex. We demonstrate room temperature condensation and propagation of polaritons in rectangular and trapezoidal waveguides by analyzing spatial and angle-resolved photoluminescence. We also discuss the blue-shifting and non-zero momentum of the condensate and show that it is strongly confined up to several higher energy levels. As an example, we report the spatial profiles of condensation in custom devices, such as a ring waveguide, a Y-splitter, and a Mach-Zehnder interferometer. This work represents a first step towards the realization of more complex, fully integrated, coherent polaritonic circuits operating at room temperature.
\end{abstract}

\section{Introduction}
Exciton-polaritons (EP) are hybrid quasiparticles consisting of mixed light-matter components, inheriting the low effective mass and strong non-linear response from their respective constituents. Realized first in semiconductor microcavities \cite{weisbuch_observation_1992}, the hybrid nature of EPs has made them highly attractive in the context of quantum nonlinear optics and classical photonic signal processing. Moreover, EPs being composite bosons, enable Bose-Einstein-like condensation \cite{imamoglu_nonequilibrium_1996,kasprzak_boseeinstein_2006,balili_bose-einstein_2007}\textemdash a process wherein the system undergoes stimulated scattering after surpassing a population threshold, leading to the formation of coherent macroscopic condensates. Since the initial observation of polariton condensation, numerous intriguing phenomena, such as superfluidity\cite{sanvitto_persistent_2010} , quantized vortices\cite{amo_superfluidity_2009} , and topological polaritonics\cite{suchomel_platform_2018} , have been demonstrated across various material platforms\cite{kavokin_polariton_2022}. Apart from interesting physics stemming from the listed phenomena, polaritons also hold great promise in quantum and classical computation by harnessing their nonlinear properties\cite{ghosh_quantum_2020,berloff_realizing_2017,dreismann_sub-femtojoule_2016,kavokin_polariton_2022}. In this context, there is much interest in realizing polaritonic circuits and related building blocks using inorganic materials\cite{wertz_spontaneous_2010,rozas_effects_2021}, perovskites\cite{su_observation_2020}, and ZnO-based systems\cite{liao_propagation_2019}. Here, we report a method for creating exciton-polariton devices on demand, such as rectangular and trapezoidal waveguides. We exhibit the emergence of a coherent polariton condensate at room temperature by analyzing both real space and angle-resolved photoluminescence (PL). We also present the real space PL of elementary components relevant for polariton circuit design, such as a ring waveguide, Y-splitter, and a Mach-Zehnder interferometer.

\begin{figure}
    \centering
    \includegraphics[width=1\linewidth]{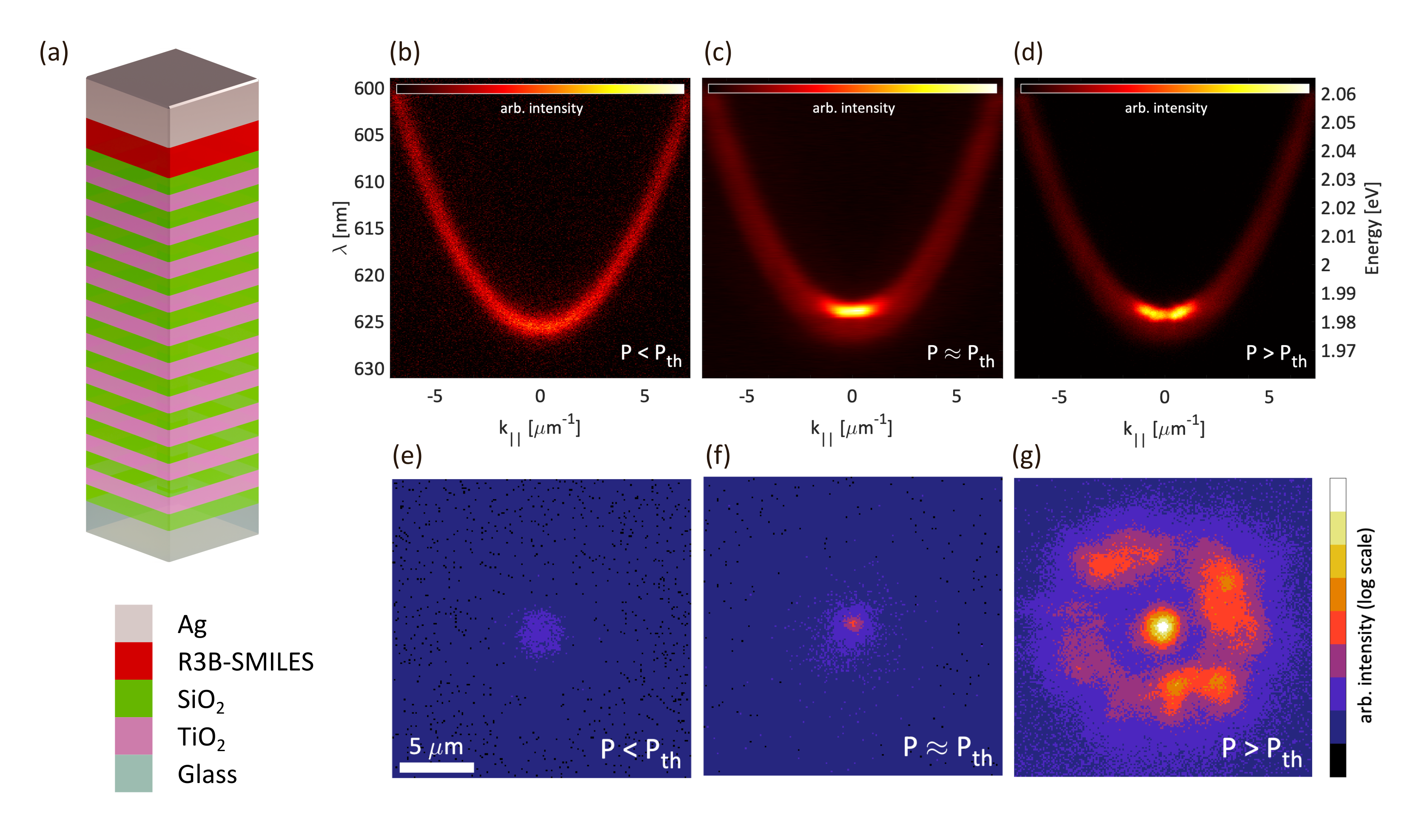}
    \caption{Cavity schematic and condensation profile. (a) Organic microcavity with active dye layer R3B-SMILES ($\sim$45 nm), bottom TiO$_2$-SiO$_2$ DBR, and top Ag mirror ($\sim$100 nm). Angle-resolved PL is shown before (b), at (c), and above (d) the condensation threshold. Corresponding real-space PL is shown in (e, f, g).}
    \label{fig:figure1}
\end{figure}

\section{Methods}
Our cavity, shown in Figure \ref{fig:figure1}(a), incorporates a distributed Bragg reflector (DBR) composed of 10.5 alternating SiO$_2$/TiO$_2$ layer pairs, centered at a center wavelength of 620 nm with a bandwidth of 200 nm. The active molecular excitonic medium, deposited on the bottom DBR via spin coating, consists of a 45-nm-thick layer of Rhodamine 3B Perchlorate (R3B) within a Small Molecule Ionic Isolation Lattices (SMILES) complex \cite{benson_plug-and-play_2020}. The Tamm plasmon cavity \cite{kar_tamm_2023} is then completed by depositing 100 nm of silver film atop the SMILES layer. These parameters produce strong exciton-photon coupling in the cavity with the lower polariton mode at 623 nm (see SI for reflectivity data). The use of R3B-SMILES as an active medium ensures a high quantum yield and enhanced photostability, which is crucial for realizing exciton-polariton condensates. Exciton-polariton condensation in the SMILES system at room temperature has previously been demonstrated\cite{deshmukh_plug-and-play_2024,choi_highly_2025} along with the realization of polaritonic lattices via FIB writing and structured illumination \cite{yadav_direct_2024,chapman_molecular_2025}.  The use of FIB writing offers the flexibility of patterning waveguides with arbitrary forms and a high refractive index contrast which confines the polaritons to the patterned devices.

We excite the microcavity samples off-resonance with a 200-femtosecond pulsed laser at a repetition rate of 0.2 kHz and a wavelength of 514 nm. We use a broad, uniform beam with a 10 µm spot size, rather than a tightly focused beam, to reduce heating-induced damage. 

\section{Results and discussion}

Emergence of polariton condensation in a planar cavity is evidenced through the collapse in momentum space to the $|\vec{k}| = 0$ state (Figure \ref{fig:figure1}(b,c)) along with localized narrow emission (Figure \ref{fig:figure1}(e,f)) as the input power approaches the threshold.  
Upon condensation, we observe a blueshift of 4 meV, which we attribute to the saturation of emitters in the organic medium at higher excitation powers \cite{betzold_coherence_2020}. This reduces the Rabi splitting between the lower and upper polariton branches and shifts the lower polariton branch to higher energies. 

As the excitation power is increased beyond the threshold, a local blueshift occurs at the pump spot, inducing a radial outflow of polaritons toward the surrounding region of lower potential. Above the threshold, we observe symmetric finite-$\vec{k}$ condensation (Figure \ref{fig:figure1}(d)) in Fourier space and an outward propagating halo of the condensate in real space (Figure \ref{fig:figure1}(g)), another confirmation that the polariton condensates are propagating in the planar region.

\begin{figure}
    \centering
    \includegraphics[width=1\linewidth]{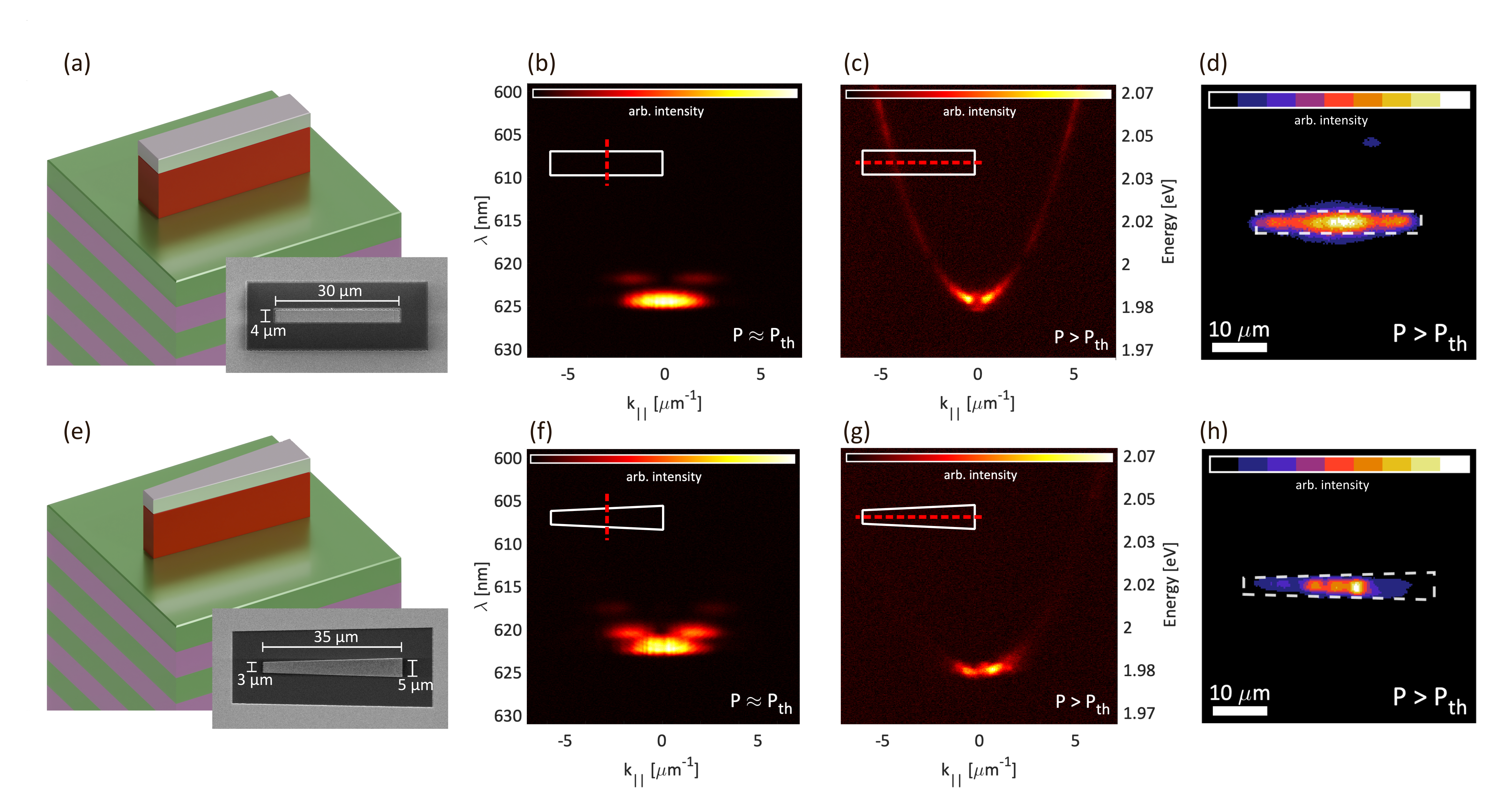}
    \caption{Finite-$\vec{k}$ condensation in rectangular and trapezoidal waveguides. The FIB-patterned rectangular wire is depicted in (a) with SEM image inset. Angle-resolved PL along the short and long axes is shown in (b) and (c), respectively. Real-space PL is shown in (d). Corresponding images for the trapezoidal case are shown in (e) through (h). Both devices were excited at the center.}
    \label{fig:figure2}
\end{figure}

After establishing condensation and propagation within the planar cavity, we fabricate additional microcavity devices into the top silver layer using a Ga\textsuperscript{+} source focused ion beam (FIB). We program this etching to create arbitrary shapes, such as a rectangular waveguide, and image them with a scanning electron microscope (SEM) (Figure \ref{fig:figure2}(a)). We mill the entire top silver layer during this process with a voltage and current of 30 kV and 0.79 nA, respectively. The patterning results in strong refractive index contrast between etched and non-etched regions. Polaritons experience this contrast as a steep potential barrier confining them within the device. Condensation within such a wire is shown in Figure \ref{fig:figure2}(d) which we will comment on further below. 

Motivated by our observation that the blue-shifting of the condensate is greater within narrower waveguides—the ground state energy is lifted for shorter confinement distances—we fabricate both rectangular (4 $\mu$m × 30 $\mu$m) and trapezoidal (3-5 $\mu$m × 35 $\mu$m) waveguides. By varying the width along the trapezoidal waveguide, we aim to induce a potential gradient encouraging the propagation of a condensate seeded at an area of higher potential. We first confirm the spatial confinement of the polariton condensates by exciting the center of the devices at the condensation threshold. We measure the angle-resolved PL of polariton condensates along the width of the waveguide. The resulting discretized energy spectra, shown in Figure \ref{fig:figure2}(b, f), confirm confinement along the width. Repeating the experiment with the slit aligned along the respective long axes, we observe symmetric finite-$\vec{k}$ condensation (Figure \ref{fig:figure2}(c)) above threshold in the rectangular waveguide, indicating symmetric polariton propagation. We also observe this in the real space image (Figure \ref{fig:figure2}(d)). In the trapezoidal waveguide, however, the so-called $``$plus-$\vec{k}$$"$ direction is favored, with the angle-resolved emission no longer symmetric with respect to $|\vec{k}|=0$ (Figure \ref{fig:figure2}(g)). This corresponds to propagation in the direction of expanding waveguide width (Figure \ref{fig:figure2}(h)). 
Here we note that the long-axis spectra only appear continuous because of the inverse-square proportionality of energy level spacing to wire length ($\Delta E \propto \frac{1}{L^2}$). For sufficiently small ($\sim$4 meV) spacings, our experimental setup is unable to resolve adjacent energy levels and the spectra appear continuous.

To verify our experimental results, we model the dynamics of the polariton wavefunction $\psi (\vec{x},t)$ with an open-dissipative Gross-Pitaevskii equation coupled to a reservoir $n_{R}(\vec{x},t)$ (see SI for more  details)
\cite{dusel_room_2020,ghosh_microcavity_2022,wouters_excitations_2007}:
\begin{subequations}
\label{eq:gpe}
\begin{equation}
\begin{split}
i \hbar \frac{\partial}{\partial t} \psi (\vec{x},t) ={}& - \frac{\hbar^2}{2m} \nabla^2 \psi (\vec{x},t) + V(\vec{x}) \psi (\vec{x},t)  + g |\psi (\vec{x},t)|^2 \psi (\vec{x},t) \\
& + g_R n_R(\vec{x},t) \psi (\vec{x},t) +  i \frac{\hbar R}{2} n_R (\vec{x},t) \psi (\vec{x},t)   -  i \frac{\hbar \gamma_c}{2} \psi (\vec{x},t)  + i \hbar \frac{\partial}{\partial t} \psi_{st} (\vec{x},t),
\end{split}
\end{equation}
\begin{equation}
\begin{split}
 \frac{\partial n_R (\vec{x},t)}{\partial t} ={}& -  \gamma_R n_R(\vec{x},t)  -   R n_R (\vec{x},t)|\psi(\vec{x},t)|^2.
\end{split}
\end{equation}
\end{subequations}

\begin{figure}
    \centering
    \includegraphics[width=1\linewidth]{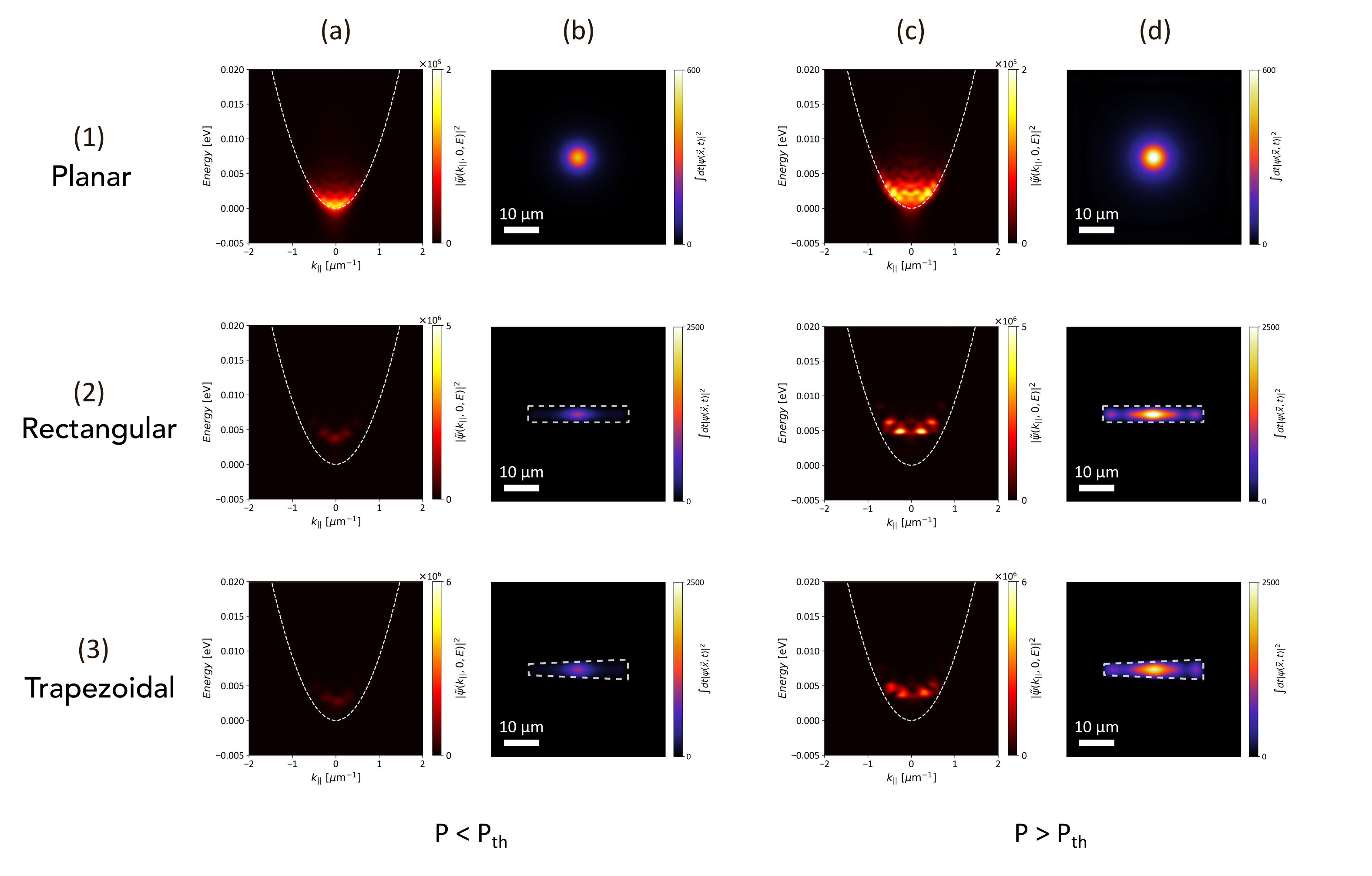}
    \caption{Numerical simulations. The GPE in Eq.~\eqref{eq:gpe} is evolved until steady state is reached, then the wavefunction is time-integrated and plotted in real space and Fourier-transformed to momentum space. The transformation is taken along the long axis where applicable. Results are shown for the planar (row 1), rectangular waveguide (row 2), and trapezoidal waveguide (row 3) cases, where the two leftmost columns are excited below condensation threshold and the two rightmost columns are excited above. The dashed white parabolas in momentum-space simulations correspond to the geometric structure (free energy dispersion). The dashed white outlines in in real-space simulations correspond to device boundaries.} 
    \label{fig:figure3}
\end{figure}

As shown in Figure \ref{fig:figure3}, the simulated and experimental data have qualitative correspondence. When the system is driven above threshold, we observe the condensate moving from $|\vec{k}|=0$ to finite-$\vec{k}$ in the simulated dispersion plots (Figure \ref{fig:figure3}(1a,1c)). A key difference, however, is the presence of multiple local maxima in simulation. This discrepancy is from the aforementioned $\sim$4 meV smearing of the experimental dispersion in Figures \ref{fig:figure1} and \ref{fig:figure2}, making the finite-$\vec{k}$ lobes appear as one global maximum. We account for this effect in the simulations by applying a Gaussian filter (see SI for more details). The real space simulations of the planar cavity (Figure \ref{fig:figure1}(b,d)) and rectangular waveguide (Figure \ref{fig:figure2}(b,d)) show the broadening of the real space pump spot and appearance of bright spots at the end of the waveguide, respectively, in good agreement with the experiment. The simulated dispersion of the trapezoidal waveguide (Figure \ref{fig:figure3}(3a, 3c)) shows asymmetry above threshold, also agreeing with experiment, though the real space results slightly differ. 

\begin{figure}
    \centering
    \includegraphics[width=0.75\linewidth]{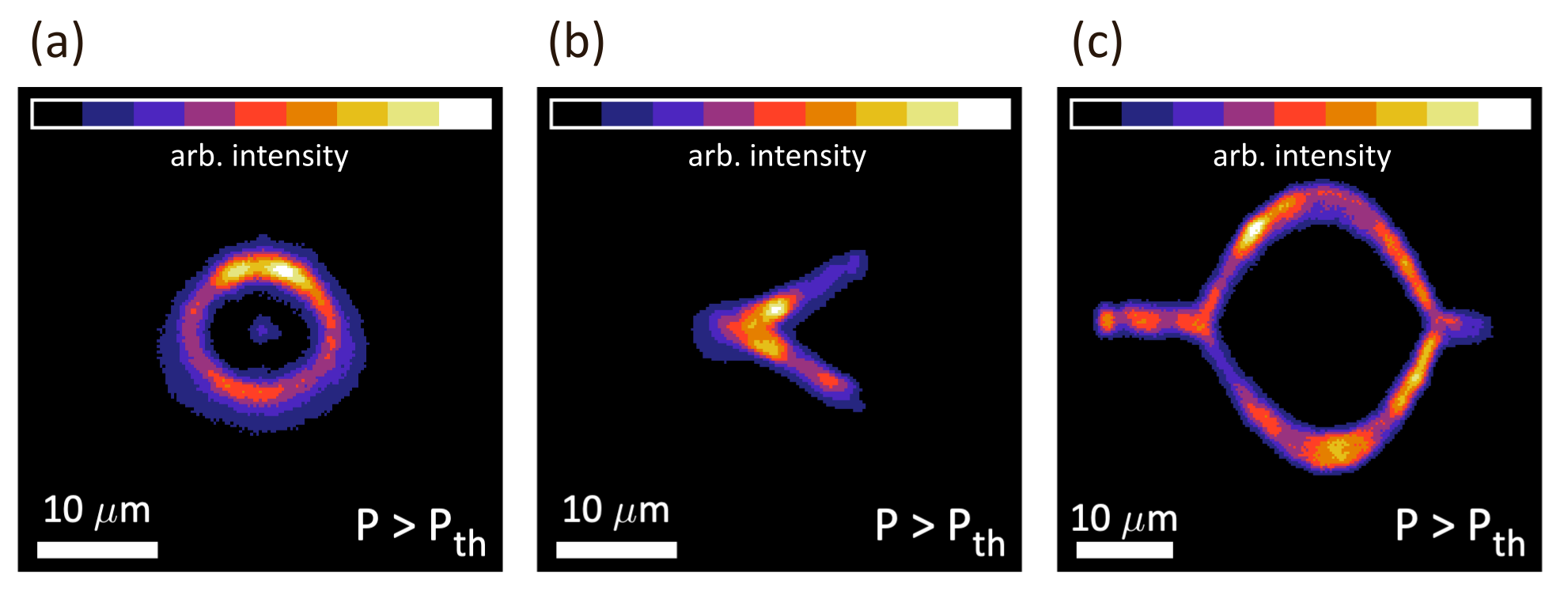}
    \caption{Experimental real-space emission above threshold from several fabricated devices: ring (a), Y-splitter (b), and Mach-Zehnder interferometer (MZI)  (c). The ring is excited with a broad ($\sim 10$ µm) laser spot, the Y-splitter with a tight ($\sim$ 2 µm) spot at the left node, and the MZI with a tight ($\sim 10$ µm) spot at the left arm.}
    \label{fig:figure4}
\end{figure}

In addition to the rectangular and trapezoidal waveguides, we fabricate a selection of other polaritonic devices as a proof of concept and to probe future research directions \cite{horowitz_long-range_2025} (Figure \ref{fig:figure4}). These include rings of diameter 5 to 25 µm, Y-splitters with arm lengths of about 10 µm, and Mach-Zehnder interferometers (MZI) \cite{kok_linear_2007} with length of about 35 µm.
 Preliminary real-space results showing polariton condensate flow are shown for these devices, further demonstrating the versatility of FIB writing and our SMILES microcavity platform to pattern any device on demand. These results pave a pathway to realize functionalities like switching at room temperature by introducing an additional laser to induce local refractive index change \cite{sturm_all-optical_2014}.  

\section{Conclusion}

In summary, we demonstrate coherent polaritonic waveguides and related device architectures operating at room temperature using molecular EP condensates. We observe strong confinement of condensates along the width of the waveguide devices as evidenced by the discrete modes in momentum space, and a propagating mode along the length of the devices characterized by finite-$\vec{k}$ condensation in $k$-space and interference fringes in real space. Ultimately, the propagation of the condensate is limited by the polariton condensate lifetime, scattering losses, and laser-heating-induced damage to the devices. For extending polariton lifetimes, there is room to increase the Q-factor of the devices \cite{choi_highly_2025} and further optimize the FIB parameters to fabricate smoother structures. On the other hand, to reduce the heating-induced damage, a thicker silver film can be deposited or an alternative top mirror can be considered. With these changes, the top-down approach of using FIB provides the flexibility to write any pattern on the cavity, thus opening possibilities to engineer complex devices in the future for building integrated coherent polariton circuits at room temperature.

\begin{acknowledgement}

The experimental work was supported by the Army Research Office grant W911NF -22-1-0091 (AS, EB, VM) and the NSF QEXPAND grant OMA-2328993 (DC, RY). The theoretical work was supported by Air Force Office of Scientific Research grant FA9550-22-1-0317. 

\end{acknowledgement}

\section{Fabrication Details}
The SMILES-based microcavity is fabricated on a quartz substrate with a distributed Bragg reflector (DBR) centered at 620 nm. The DBR, composed of 10.5 pairs of SiO$_2$/TiO$_2$, was cleaned using O$_2$ plasma for 5 min. A 45 nm thick SMILES film is then deposited on the cleaned DBR using a two-step spin coating technique. In the first step, the film is spun for 20 s at a speed of 1000 rpm, followed by a a second spin for 80 s at 3000 rpm. The deposited SMILES film is placed in a constant pressure vacuum at 25 °C. Finally, a 100 nm thick silver layer is deposited atop the SMILES film using an e-beam evaporator, completing the fabrication of the microcavity.

\section{Experimental Details}

\begin{figure}
    \centering
    \includegraphics[width=0.75\linewidth]{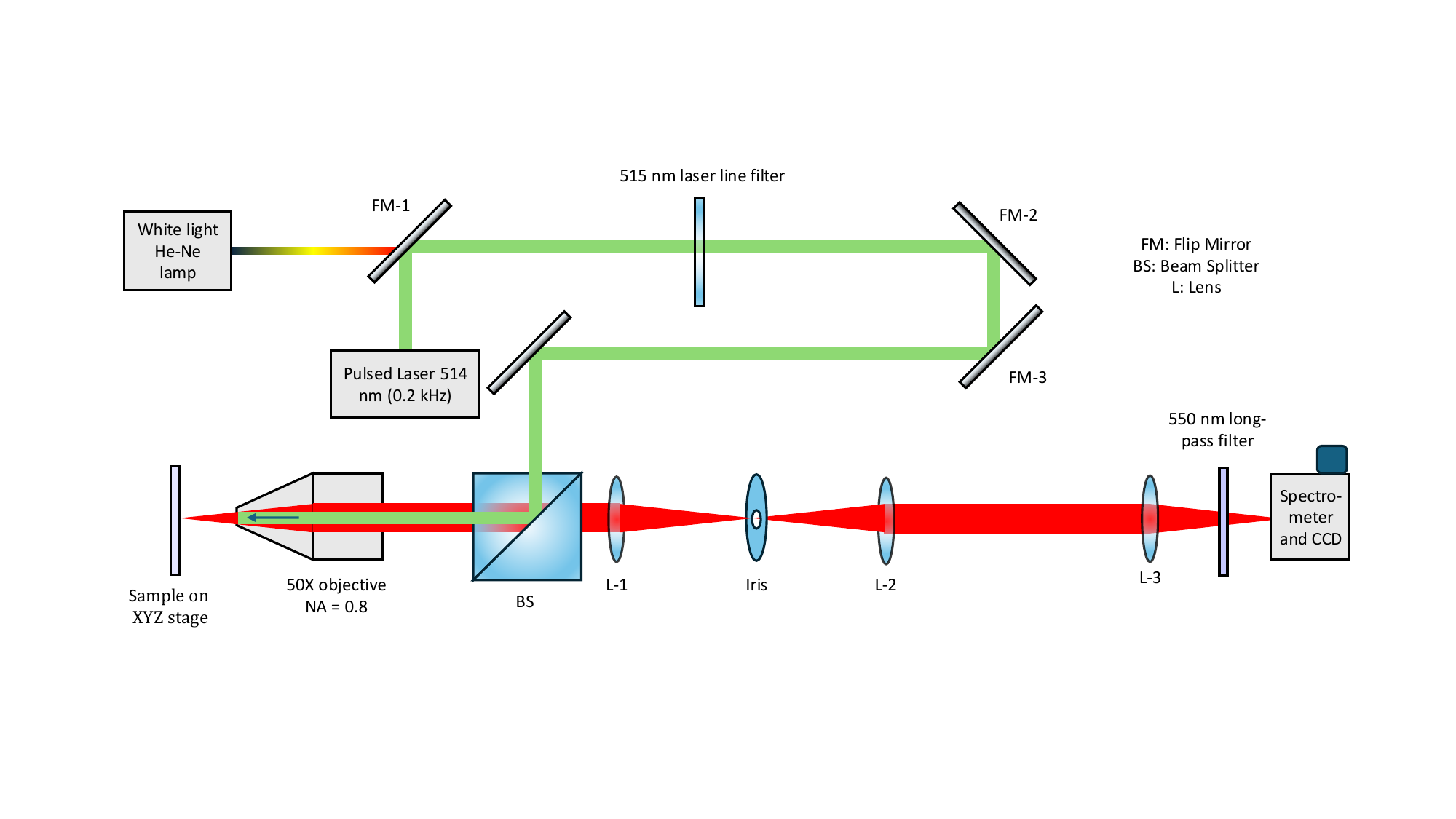}
    \caption{Schematic of experimental optical setup.}
    \label{fig:figure5}
\end{figure}

We employ real space and Fourier space PL imaging to map the condensate emission characteristics. Energy-resolved angle-dependent PL is measured by selecting a specific in-plane momentum using a monochromator (Princeton Instruments) in front of the CCD camera and dispersing the PL using a 300 g/mm grating. We use a pulsed 514 nm laser (Carbide from Light Conversion) with a pulse width of 280 fs and a repetition rate of 0.2 kHz for the condensation experiments. A 550 nm long-pass filter is placed before the CCD to clean laser light from measurements. A schematic of the experimental setup is presented in Figure \ref{fig:figure5}.

\section{Data Availability Statement}
All data will be provided by the corresponding authors upon reasonable request.

\bibliography{mainreferences}

@article{chapman_molecular_2025,
	title = {Molecular exciton-polariton condensate lattice via structured illumination},
	volume = {12},
	copyright = {© 2025 Optica Publishing Group},
	issn = {2334-2536},
	url = {https://opg.optica.org/optica/abstract.cfm?uri=optica-12-12-1873},
	doi = {10.1364/OPTICA.577660},
	language = {EN},
	number = {12},
	urldate = {2025-12-01},
	journal = {Optica},
	author = {Chapman, Jackson and Choi, Daegwang and Yadav, Ravindra Kumar and Bader, Ezra and Cognee, Kevin G. and Satapathy, Sitakanta and Menon, Vinod M.},
	month = dec,
	year = {2025},
	note = {Publisher: Optica Publishing Group},
	pages = {1873--1877},
}

@article{sturm_all-optical_2014,
	title = {All-optical phase modulation in a cavity-polariton {Mach}–{Zehnder} interferometer},
	volume = {5},
	copyright = {2014 The Author(s)},
	issn = {2041-1723},
	url = {https://www.nature.com/articles/ncomms4278},
	doi = {10.1038/ncomms4278},
	language = {en},
	number = {1},
	urldate = {2025-12-01},
	journal = {Nature Communications},
	author = {Sturm, C. and Tanese, D. and Nguyen, H. S. and Flayac, H. and Galopin, E. and Lemaître, A. and Sagnes, I. and Solnyshkov, D. and Amo, A. and Malpuech, G. and Bloch, J.},
	month = feb,
	year = {2014},
	note = {Publisher: Nature Publishing Group},
	pages = {3278},
}

@article{kok_linear_2007,
	title = {Linear optical quantum computing with photonic qubits},
	volume = {79},
	url = {https://link.aps.org/doi/10.1103/RevModPhys.79.135},
	doi = {10.1103/RevModPhys.79.135},
	number = {1},
	urldate = {2025-11-12},
	journal = {Reviews of Modern Physics},
	author = {Kok, Pieter and Munro, W. J. and Nemoto, Kae and Ralph, T. C. and Dowling, Jonathan P. and Milburn, G. J.},
	month = jan,
	year = {2007},
	note = {Publisher: American Physical Society},
	pages = {135--174},
}

@article{kar_tamm_2023,
	title = {Tamm plasmon polariton in planar structures: {A} brief overview and applications},
	volume = {159},
	issn = {0030-3992},
	shorttitle = {Tamm plasmon polariton in planar structures},
	url = {https://www.sciencedirect.com/science/article/pii/S003039922201074X},
	doi = {10.1016/j.optlastec.2022.108928},
	urldate = {2025-11-12},
	journal = {Optics \& Laser Technology},
	author = {Kar, Chinmaya and Jena, Shuvendu and Udupa, Dinesh V. and Rao, K. Divakar},
	month = apr,
	year = {2023},
	pages = {108928},
}

@article{choi_highly_2025,
	title = {Highly {Coherent} {Room}-temperature {Molecular} {Polariton} {Condensates}},
	volume = {13},
	copyright = {© 2025 Wiley-VCH GmbH},
	issn = {2195-1071},
	url = {https://onlinelibrary.wiley.com/doi/abs/10.1002/adom.202500086},
	doi = {10.1002/adom.202500086},
	language = {en},
	number = {16},
	urldate = {2025-11-09},
	journal = {Advanced Optical Materials},
	author = {Choi, Daegwang and Zachariah, Serena and Yadav, Ravindra Kumar and Menon, Vinod. M.},
	year = {2025},
	note = {\_eprint: https://advanced.onlinelibrary.wiley.com/doi/pdf/10.1002/adom.202500086},
	pages = {2500086},
}

@article{horowitz_long-range_2025,
	title = {Long-{Range} {Coherent} {Emission} of {Propagating} {Exciton}–{Polaritons} from a {Mach}–{Zehnder} {Interferometer}},
	volume = {12},
	url = {https://doi.org/10.1021/acsphotonics.4c02442},
	doi = {10.1021/acsphotonics.4c02442},
	number = {4},
	urldate = {2025-11-09},
	journal = {ACS Photonics},
	author = {Horowitz, Jeffrey and Liu, Bin and Paul, Sritoma and Forrest, Stephen R.},
	month = apr,
	year = {2025},
	note = {Publisher: American Chemical Society},
	pages = {2027--2033},
}

@article{weisbuch_observation_1992,
	title = {Observation of the coupled exciton-photon mode splitting in a semiconductor quantum microcavity},
	volume = {69},
	url = {https://link.aps.org/doi/10.1103/PhysRevLett.69.3314},
	doi = {10.1103/PhysRevLett.69.3314},
	number = {23},
	urldate = {2025-11-09},
	journal = {Physical Review Letters},
	author = {Weisbuch, C. and Nishioka, M. and Ishikawa, A. and Arakawa, Y.},
	month = dec,
	year = {1992},
	note = {Publisher: American Physical Society},
	pages = {3314--3317},
}

@article{kavokin_polariton_2022,
	title = {Polariton condensates for classical and quantum computing},
	volume = {4},
	issn = {2522-5820},
	url = {https://www.nature.com/articles/s42254-022-00447-1},
	doi = {10.1038/s42254-022-00447-1},
	language = {en},
	number = {7},
	urldate = {2025-11-09},
	journal = {Nature Reviews Physics},
	author = {Kavokin, Alexey and Liew, Timothy C. H. and Schneider, Christian and Lagoudakis, Pavlos G. and Klembt, Sebastian and Hoefling, Sven},
	month = apr,
	year = {2022},
	pages = {435--451},
}

@article{ghosh_microcavity_2022,
	title = {Microcavity exciton polaritons at room temperature},
	volume = {1},
	issn = {2791-1748, 2791-1748},
	url = {https://www.spiedigitallibrary.org/journals/photonics-insights/volume-1/issue-1/R04/Microcavity-exciton-polaritons-at-room-temperature/10.3788/PI.2022.R04.full},
	doi = {10.3788/PI.2022.R04},
	number = {1},
	urldate = {2025-10-30},
	journal = {Photonics Insights},
	author = {Ghosh, Sanjib and Su, Rui and Zhao, Jiaxin and Fieramosca, Antonio and Wu, Jinqi and Li, Tengfei and Zhang, Qing and Li, Feng and Chen, Zhanghai and Liew, Timothy and Sanvitto, Daniele and Xiong, Qihua},
	month = aug,
	year = {2022},
	note = {Publisher: SPIE},
	pages = {R04},
}

@article{wouters_excitations_2007,
	title = {Excitations in a {Nonequilibrium} {Bose}-{Einstein} {Condensate} of {Exciton} {Polaritons}},
	volume = {99},
	doi = {10.1103/PhysRevLett.99.140402},
	number = {14},
	journal = {Physical Review Letters},
	author = {Wouters, Michiel},
	year = {2007},
}

@article{dusel_room_2020,
	title = {Room temperature organic exciton–polariton condensate in a lattice},
	volume = {11},
	copyright = {2020 The Author(s)},
	issn = {2041-1723},
	url = {https://www.nature.com/articles/s41467-020-16656-0},
	doi = {10.1038/s41467-020-16656-0},
	language = {en},
	number = {1},
	urldate = {2025-10-30},
	journal = {Nature Communications},
	author = {Dusel, M. and Betzold, S. and Egorov, O. A. and Klembt, S. and Ohmer, J. and Fischer, U. and Höfling, S. and Schneider, C.},
	month = jun,
	year = {2020},
	note = {Publisher: Nature Publishing Group},
	pages = {2863},
}

@article{dreismann_sub-femtojoule_2016,
	title = {A sub-femtojoule electrical spin-switch based on optically trapped polariton condensates},
	volume = {15},
	issn = {1476-1122, 1476-4660},
	url = {https://www.nature.com/articles/nmat4722},
	doi = {10.1038/nmat4722},
	language = {en},
	number = {10},
	urldate = {2025-10-20},
	journal = {Nature Materials},
	author = {Dreismann, Alexander and Ohadi, Hamid and Del Valle-Inclan Redondo, Yago and Balili, Ryan and Rubo, Yuri G. and Tsintzos, Simeon I. and Deligeorgis, George and Hatzopoulos, Zacharias and Savvidis, Pavlos G. and Baumberg, Jeremy J.},
	month = oct,
	year = {2016},
	pages = {1074--1078},
}

@article{betzold_coherence_2020,
	title = {Coherence and {Interaction} in {Confined} {Room}-{Temperature} {Polariton} {Condensates} with {Frenkel} {Excitons}},
	volume = {7},
	url = {https://doi.org/10.1021/acsphotonics.9b01300},
	doi = {10.1021/acsphotonics.9b01300},
	number = {2},
	urldate = {2025-10-20},
	journal = {ACS Photonics},
	author = {Betzold, Simon and Dusel, Marco and Kyriienko, Oleksandr and Dietrich, Christof P. and Klembt, Sebastian and Ohmer, Jürgen and Fischer, Utz and Shelykh, Ivan A. and Schneider, Christian and Höfling, Sven},
	month = feb,
	year = {2020},
	note = {Publisher: American Chemical Society},
	pages = {384--392},
}

@article{berloff_realizing_2017,
	title = {Realizing the \${XY}\$ {Hamiltonian} in polariton simulators},
	volume = {16},
	issn = {1476-1122, 1476-4660},
	url = {http://arxiv.org/abs/1607.06065},
	doi = {10.1038/nmat4971},
	number = {11},
	urldate = {2025-10-20},
	journal = {Nature Materials},
	author = {Berloff, Natalia G. and Kalinin, Kirill and Silva, Matteo and Langbein, Wolfgang and Lagoudakis, Pavlos G.},
	month = nov,
	year = {2017},
	note = {arXiv:1607.06065 [cond-mat]},
	pages = {1120--1126},
}

@article{benson_plug-and-play_2020,
	title = {Plug-and-{Play} {Optical} {Materials} from {Fluorescent} {Dyes} and {Macrocycles}},
	volume = {6},
	issn = {2451-9294},
	url = {https://www.sciencedirect.com/science/article/pii/S2451929420303107},
	doi = {10.1016/j.chempr.2020.06.029},
	number = {8},
	urldate = {2025-10-20},
	journal = {Chem},
	author = {Benson, Christopher R. and Kacenauskaite, Laura and VanDenburgh, Katherine L. and Zhao, Wei and Qiao, Bo and Sadhukhan, Tumpa and Pink, Maren and Chen, Junsheng and Borgi, Sina and Chen, Chun-Hsing and Davis, Brad J. and Simon, Yoan C. and Raghavachari, Krishnan and Laursen, Bo W. and Flood, Amar H.},
	month = aug,
	year = {2020},
	pages = {1978--1997},
}

@article{deshmukh_plug-and-play_2024,
	title = {Plug-and-{Play} {Molecular} {Approach} for {Room} {Temperature} {Polariton} {Condensation}},
	volume = {11},
	url = {https://doi.org/10.1021/acsphotonics.3c01547},
	doi = {10.1021/acsphotonics.3c01547},
	number = {2},
	urldate = {2025-10-20},
	journal = {ACS Photonics},
	author = {Deshmukh, Prathmesh and Satapathy, Sitakanta and Michail, Evripidis and Olsson, Andrew H. and Bushati, Rezlind and Yadav, Ravindra Kumar and Khatoniar, Mandeep and Chen, Junsheng and John, George and Laursen, Bo W. and Flood, Amar H. and Sfeir, Matthew Y. and Menon, Vinod M.},
	month = feb,
	year = {2024},
	note = {Publisher: American Chemical Society},
	pages = {348--355},
}

@article{yadav_direct_2024,
	title = {Direct {Writing} of {Room} {Temperature} {Polariton} {Condensate} {Lattice}},
	volume = {24},
	issn = {1530-6984},
	url = {https://doi.org/10.1021/acs.nanolett.4c00586},
	doi = {10.1021/acs.nanolett.4c00586},
	number = {16},
	urldate = {2025-10-20},
	journal = {Nano Letters},
	author = {Yadav, Ravindra Kumar and Satapathy, Sitakanta and Deshmukh, Prathmesh and Datta, Biswajit and Sharma, Addhyaya and Olsson, Andrew H. and Chen, Junsheng and Laursen, Bo W. and Flood, Amar H. and Sfeir, Matthew Y. and Menon, Vinod M.},
	month = apr,
	year = {2024},
	note = {Publisher: American Chemical Society},
	pages = {4945--4950},
}

@article{liao_propagation_2019,
	title = {Propagation of a polariton condensate in a one-dimensional microwire at room temperature},
	volume = {12},
	issn = {1882-0786},
	url = {https://doi.org/10.7567/1882-0786/ab1186},
	doi = {10.7567/1882-0786/ab1186},
	language = {en},
	number = {5},
	urldate = {2025-10-20},
	journal = {Applied Physics Express},
	author = {Liao, Liming and Ling, Yanjing and Luo, Song and Zhang, Zhe and Wang, Jun and Chen, Zhanghai},
	month = apr,
	year = {2019},
	note = {Publisher: IOP Publishing},
	pages = {052009},
}

@article{rozas_effects_2021,
	title = {Effects of the {Linear} {Polarization} of {Polariton} {Condensates} in {Their} {Propagation} in {Codirectional} {Couplers}},
	volume = {8},
	url = {https://doi.org/10.1021/acsphotonics.1c00746},
	doi = {10.1021/acsphotonics.1c00746},
	number = {8},
	urldate = {2025-10-20},
	journal = {ACS Photonics},
	author = {Rozas, Elena and Yulin, Alexey and Beierlein, Johannes and Klembt, Sebastian and Höfling, Sven and Egorov, Oleg and Peschel, Ulf and Shelykh, I. A. and Gundin, Manuel and Robles-López, Ignacio and Martín, M. D. and Viña, L.},
	month = aug,
	year = {2021},
	note = {Publisher: American Chemical Society},
	pages = {2489--2497},
}

@article{su_observation_2020,
	title = {Observation of exciton polariton condensation in a perovskite lattice at room temperature},
	volume = {16},
	copyright = {2020 The Author(s), under exclusive licence to Springer Nature Limited},
	issn = {1745-2481},
	url = {https://www.nature.com/articles/s41567-019-0764-5},
	doi = {10.1038/s41567-019-0764-5},
	language = {en},
	number = {3},
	urldate = {2025-10-20},
	journal = {Nature Physics},
	author = {Su, Rui and Ghosh, Sanjib and Wang, Jun and Liu, Sheng and Diederichs, Carole and Liew, Timothy C. H. and Xiong, Qihua},
	month = mar,
	year = {2020},
	note = {Publisher: Nature Publishing Group},
	pages = {301--306},
}

@article{suchomel_platform_2018,
	title = {Platform for {Electrically} {Pumped} {Polariton} {Simulators} and {Topological} {Lasers}},
	volume = {121},
	issn = {0031-9007, 1079-7114},
	url = {https://link.aps.org/doi/10.1103/PhysRevLett.121.257402},
	doi = {10.1103/PhysRevLett.121.257402},
	language = {en},
	number = {25},
	urldate = {2025-10-20},
	journal = {Physical Review Letters},
	author = {Suchomel, Holger and Klembt, Sebastian and Harder, Tristan H. and Klaas, Martin and Egorov, Oleg A. and Winkler, Karol and Emmerling, Monika and Thomale, Ronny and Höfling, Sven and Schneider, Christian},
	month = dec,
	year = {2018},
	pages = {257402},
}

@article{wertz_spontaneous_2010,
	title = {Spontaneous formation and optical manipulation of extended polariton condensates},
	volume = {6},
	copyright = {2010 Springer Nature Limited},
	issn = {1745-2481},
	url = {https://www.nature.com/articles/nphys1750},
	doi = {10.1038/nphys1750},
	language = {en},
	number = {11},
	urldate = {2025-10-20},
	journal = {Nature Physics},
	author = {Wertz, E. and Ferrier, L. and Solnyshkov, D. D. and Johne, R. and Sanvitto, D. and Lemaître, A. and Sagnes, I. and Grousson, R. and Kavokin, A. V. and Senellart, P. and Malpuech, G. and Bloch, J.},
	month = nov,
	year = {2010},
	note = {Publisher: Nature Publishing Group},
	pages = {860--864},
}

@article{sanvitto_persistent_2010,
	title = {Persistent currents and quantized vortices in a polariton superfluid},
	volume = {6},
	copyright = {2010 Springer Nature Limited},
	issn = {1745-2481},
	url = {https://www.nature.com/articles/nphys1668},
	doi = {10.1038/nphys1668},
	language = {en},
	number = {7},
	urldate = {2025-10-20},
	journal = {Nature Physics},
	author = {Sanvitto, D. and Marchetti, F. M. and Szymańska, M. H. and Tosi, G. and Baudisch, M. and Laussy, F. P. and Krizhanovskii, D. N. and Skolnick, M. S. and Marrucci, L. and Lemaître, A. and Bloch, J. and Tejedor, C. and Viña, L.},
	month = jul,
	year = {2010},
	note = {Publisher: Nature Publishing Group},
	pages = {527--533},
}

@article{amo_superfluidity_2009,
	title = {Superfluidity of polaritons in semiconductor microcavities},
	volume = {5},
	copyright = {2009 Springer Nature Limited},
	issn = {1745-2481},
	url = {https://www.nature.com/articles/nphys1364},
	doi = {10.1038/nphys1364},
	language = {en},
	number = {11},
	urldate = {2025-10-20},
	journal = {Nature Physics},
	author = {Amo, Alberto and Lefrère, Jérôme and Pigeon, Simon and Adrados, Claire and Ciuti, Cristiano and Carusotto, Iacopo and Houdré, Romuald and Giacobino, Elisabeth and Bramati, Alberto},
	month = nov,
	year = {2009},
	note = {Publisher: Nature Publishing Group},
	pages = {805--810},
}

@article{imamoglu_nonequilibrium_1996,
	title = {Nonequilibrium condensates and lasers without inversion: {Exciton}-polariton lasers},
	volume = {53},
	shorttitle = {Nonequilibrium condensates and lasers without inversion},
	doi = {10.1103/PhysRevA.53.4250},
	number = {6},
	journal = {Physical Review A},
	author = {Imamoglu, A.},
	year = {1996},
	pages = {4250--4253},
}

@article{balili_bose-einstein_2007,
	title = {Bose-{Einstein} {Condensation} of {Microcavity} {Polaritons} in a {Trap}},
	volume = {316},
	issn = {0036-8075, 1095-9203},
	url = {https://www.science.org/doi/10.1126/science.1140990},
	doi = {10.1126/science.1140990},
	language = {en},
	number = {5827},
	urldate = {2025-10-20},
	journal = {Science},
	author = {Balili, R. and Hartwell, V. and Snoke, D. and Pfeiffer, L. and West, K.},
	month = may,
	year = {2007},
	pages = {1007--1010},
}

@article{kasprzak_boseeinstein_2006,
	title = {Bose–{Einstein} condensation of exciton polaritons},
	volume = {443},
	copyright = {2006 Springer Nature Limited},
	issn = {1476-4687},
	url = {https://www.nature.com/articles/nature05131},
	doi = {10.1038/nature05131},
	language = {en},
	number = {7110},
	urldate = {2025-10-20},
	journal = {Nature},
	author = {Kasprzak, J. and Richard, M. and Kundermann, S. and Baas, A. and Jeambrun, P. and Keeling, J. M. J. and Marchetti, F. M. and Szymańska, M. H. and André, R. and Staehli, J. L. and Savona, V. and Littlewood, P. B. and Deveaud, B. and Dang, Le Si},
	month = sep,
	year = {2006},
	note = {Publisher: Nature Publishing Group},
	pages = {409--414},
}

@article{ghosh_quantum_2020,
	title = {Quantum computing with exciton-polariton condensates},
	volume = {6},
	copyright = {2020 The Author(s)},
	issn = {2056-6387},
	url = {https://www.nature.com/articles/s41534-020-0244-x},
	doi = {10.1038/s41534-020-0244-x},
	language = {en},
	number = {1},
	urldate = {2025-10-20},
	journal = {npj Quantum Information},
	author = {Ghosh, Sanjib and Liew, Timothy C. H.},
	month = feb,
	year = {2020},
	note = {Publisher: Nature Publishing Group},
	pages = {16},
}
\end{document}